\documentclass{article}
\usepackage{authblk}
\usepackage{amsmath,amsfonts,amssymb}
\usepackage{txfonts}
\usepackage[T1]{fontenc}
\usepackage{hyperref}

\def\bd{\begin{displaymath}}\def\ed{\end{displaymath}}
\def\be{\begin{equation}}\def\ee{\end{equation}}
\def\bea{\begin{eqnarray}}\def\eea{\end{eqnarray}}
\def\ba{\begin{array}}\def\ea{\end{array}}

\begin{document}

\begin{titlepage}
%\pagenumbering{gobble}
\title{Interpolations between Jordanian twists, the Poincar\'e-Weyl algebra and dispersion relations}
\author[1]{D. Meljanac{\footnote{Daniel.Meljanac@irb.hr}}}
\author[2]{S. Meljanac{\footnote{meljanac@irb.hr}}}
\author[3]{Z. \v{S}koda{\footnote{zoran.skoda@uhk.cz}}}
\author[4]{R. \v Strajn{\footnote{rina.strajn@unidu.hr}}}
\affil[1]{Division of Materials Physics, Ru\dj er Bo\v skovi\'c Institute, Bijeni\v cka cesta 54, 10002 Zagreb, Croatia}
\affil[2]{Division of Theoretical Physics, Ru\dj er Bo\v skovi\'c Institute, Bijeni\v cka cesta 54, 10002 Zagreb, Croatia}
\affil[3]{Faculty of Science, University of  Hradec Kr\'{a}lov\'{e}, Rokitansk\'{e}ho 62, Hradec Kr\'{a}lov\'{e}, Czech Republic}
\affil[4]{Department of Electrical Engineering and Computing, University of Dubrovnik, \'{C}ira Cari\'{c}a 4, 20000 Dubrovnik, Croatia}
\date{}
\renewcommand\Affilfont{\itshape\small}
\clearpage\maketitle
\thispagestyle{empty}

\begin{abstract}
We consider a two parameter family of Drinfeld twists generated from a simple Jordanian twist further twisted by 1-cochains. Twists from this family interpolate between two simple Jordanian twists. Relations between them are constructed and discussed. It is proved that there exists a one parameter family of twists identical to a simple Jordanian twist. The twisted coalgebra, star product and coordinate realizations of the $\kappa$-Minkowski noncommutative space time are presented. Real forms of Jordanian deformations are also discussed. The method of similarity transformations is applied to the Poincar\'e-Weyl Hopf algebra and two types of one parameter families of dispersion relations are constructed. Mathematically equivalent deformations, that are related to nonlinear changes of symmetry generators and linked with similarity maps, may lead to differences in the description of physical phenomena.

\bigskip

\noindent \textit{keywords}: noncommutative geometry; twist deformation; $\kappa$-Minkowski spacetime; Poincar\'e-Weyl algebra; dispersion relations

\bigskip

% Quantum groups, noncommutative geometry, dispersion relations

\end{abstract}

\end{titlepage}

\section{Introduction}

Many proposals to resolve fundamental issues at the Planck scale involve the development of models of field theories on noncommutative (NC) spaces, most notably the $\kappa$-Minkowski spacetime. The parameter $\kappa$ is here usually interpreted as the Planck mass or the quantum gravity scale. One of the possible quantum symmetries of the $\kappa$-Minkowski NC space time is the $\kappa$-Poincar\'e quantum group \cite{LukTol,LukRuegg} and it constitutes one of the examples of deformed relativistic spacetime symmetries and the corresponding dispersion relations. Similar models exhibit Hopf algebra symmetries.

For Hopf algebras there is a remarkable systematic twisting procedure, invented by Drinfeld \cite{drinfeld,majid}. Namely, from a given Hopf algebra $H$ and a twist element $\mathcal{F}\in H \otimes H$ satisfying the 2-cocycle condition, one produces a new Hopf algebra $H^{\mathcal{F}}$ with the same algebra sector, but a different coalgebra sector, with coproduct $\Delta^{\mathcal{F}}(h)=\mathcal{F}\Delta (h) \mathcal{F}^{-1}$. If a spacetime has a Hopf algebra symmetry, and the Hopf algebra is twisted, then the spacetime can also be twisted using the same twist and preserving the covariance properties. Moreover, many other constructions like differential calculi and, to some extent, some basic constructions of field theories can be systematically deformed by procedures involving only a twist. Thus, the Drinfeld twist provides a laboratory for systematic deformation of space time and for investigating its deformed relativistic symmetry, geometric and physical structures.

Drinfeld 2-cocycles can be further modified by 1-cochains \cite{drinfeld,majid}. Given a Drinfeld twist $\mathcal{F}$ and an invertible 1-cochain $\omega \in H$, the expression $\mathcal{F}_\omega = (\omega^{-1} \otimes \omega^{-1}) \mathcal{F} \Delta (\omega)$ defines a new Drinfeld twist cohomologous to $\mathcal{F}$. If we start from a trivial 2-cocycle $1\otimes 1$, we obtain a twist, $(\omega^{-1} \otimes \omega^{-1}) \Delta (\omega)$, which is a 2-coboundary in the sense of nonabelian cohomology \cite{majid}. Cohomologous 2-cocycles induce isomorphic deformed Hopf algebras and equivalent related mathematical constructions. Hence, we may say that the transformation of changing a 2-cocycle by a 1-cochain is a gauge transformation of the 2-cocycle.

In the late 1980-s remarkable deformations of $R$-matrices and related quantum groups have been found under the name of Jordan(ian) $R$-matrices and Jordanian deformations \cite{lyub,demidov}. The corresponding Drinfeld twist has been written out independently in Refs. \cite{CollGerstGiaq1989} and \cite{Og}. These examples involve the universal enveloping algebra of the two dimensional solvable Lie algebra (with generators $H,\,E,\, [H,E] = E$), some Hopf algebra which contains it ($U(\mathfrak{sl}(2))$, Yangian $Y(\mathfrak{gl}(2))$ etc.) or their duals. A special example of a Jordanian twist is $\mathcal{F}_0= \text{exp} (\ln (1+\alpha E) \otimes H) =1\otimes 1 +\alpha E\otimes H + O(\alpha^2)$ with $\alpha$ a deformation parameter \cite{Og}. Also, $r$-symmetric versions of the Jordanian twist, where $r= E\otimes H -H\otimes E$ is the classical $r$-matrix, were introduced in Refs. \cite{Tolstoy,Tolstoy2} and \cite{GZ}.

The Jordanian twist reappeared in the context of the $\kappa$-Minkowski NC space time \cite{BP,BuKim,achieriobs,mmmsprd}. A relation with the symmetry of the $\kappa$-Minkowski space time is established with the introduction of the generators of relativistic symmetries, dilatation $D$ and momenta $p_\alpha$ (instead of generators $H$ and $E$ of the $\mathfrak{sl}(2)$ algebra) satisfying the same commutation relation, $[p_\alpha,D ] = p_\alpha$. Dilatation $D$ is included in a minimal extension of the relativistic space time symmetry, the so-called Poincar\'e-Weyl algebra generated by $\{ M_{\mu\nu},\, p_\mu,\, D\}$, where $M_{\mu\nu}$ denote the Lorentz generators. One parameter interpolations between Jordanian twists, which are generated from a simple Jordanian twist $\mathcal{F}_0$ by twisting with 1-cochains, were studied in Refs. \cite{remarks,cobtw,MMss}.
 
Applications of Jordanian twists have been of interest in recent literature. For example, Jordanian deformations of the conformal algebra were considered in Refs.~\cite{meljpacholpikutic}, \cite{ballesteros}, \cite{b2} and \cite{b3}. In Refs. \cite{ballesteros,b2,b3}, deformations of the anti de Sitter and the de Sitter algebra were investigated. Jordanian deformations have also been considered within applications in the $AdS$/CFT correspondence \cite{kmy,kmy2,kmy3,kmy4}. Integrable deformations of sigma models in relation to deformations of $AdS_5$ and supergravity were investigated \cite{tongeren,tong2}. Jordanian twists have been applied in the deformation of spacetime metrics \cite{borowtetrad}, dispersion relations \cite{achieriobs,mmmsprd} and gauge theories \cite{dimit}.

In this paper, we consider a two parameter family of Drinfeld twists generated from a simple Jordanian twist by further twisting by 1-cochains. This is a generalization of results presented in Refs. \cite{remarks,cobtw,MMss}. The first part of the paper is a sort of supplement to Ref. \cite{cobtw} with new results presented in Sections 3, 4 and 5. This two parameter generalization leads to the $\kappa$-Minkowski spacetime and produces the same deformation of the Poincar\'e-Weyl symmetry algebra. We show that twists from this family interpolate between two simple Jordanian twists. Using the method of similarity transformations, we construct one parameter families of dispersion relations related to each other by inverse transformation. We point out that mathematically equivalent deformations, that are related to nonlinear changes of the symmetry generators and linked with similarity maps, may lead to differences in the description of physical phenomena.

In Section 2, two special families of twists induced by 1-cochains are presented with results rewritten from Refs. \cite{cobtw,MMss}. These twists interpolate between two simple Jordanian twists. In Section 3, we define a two parameter family of Jordanian twists which is a generalization from Ref. \cite{cobtw}. Relations between them are presented and discussed. In Section 4 we also generalize results from Ref. \cite{cobtw}. In 4.1, the twisted coalgebra sector and, in 4.2, coordinate realizations and the star product are presented.
Real forms of new Jordanian deformations are discussed in Subsection 4.3. In Section 5, a method of similarity transformations is applied to the Poincar\'e-Weyl Hopf algebra and one parameter families of dispersion relations are constructed. At the end of Section 5, concluding remarks are given.

\section{Families $\mathcal{F}_{L,u}$ and $\mathcal{F}_{R,u}$ of Jordanian twists and the relation between them}

In Ref. \cite{remarks}, the following family of Drinfeld twists was considered
\begin{eqnarray}
\mathcal{F}_{L,u}&=&\text{exp}\Big( -u\left( DA\otimes 1+1 \otimes DA\right) \Big)\, \text{exp}\Big( -\text{ln}(1+A)\otimes D\Big)\, \text{exp}\Big(\Delta (uDA)\Big)\\
&=& \text{exp}\left( \frac{u}{\kappa}\left( DP\otimes 1+1 \otimes DP\right) \right)\, \text{exp}\left( -\text{ln}\left(1-\frac{1}{\kappa}P\right)\otimes D\right) \, \text{exp}\left(-\Delta \left(\frac{u}{\kappa} DP\right)\right),\label{FLu2}
\end{eqnarray}
where generators of dilatation $D$ and momenta $P$ satisfy $[P,D]=P$ and $A=-\frac{1}{\kappa}P$. The deformation parameter $\kappa$ is of the order of the Planck mass, $u$ is a real dimensionless parameter and $\Delta$ is the undeformed coproduct. These twists are constructed using a 1-cochain $\omega_L=\text{exp}\left(-\frac{u}{\kappa}DP\right)$ and they satisfy the normalization and cocycle condition.

The corresponding deformed Hopf algebra is given by
\begin{eqnarray}
&& \Delta^{\mathcal{F}_{L,u}} (p_\mu) =\mathcal{F}_{L,u}\, \Delta p_\mu\, \mathcal{F}_{L,u}^{-1} =\frac{p_\mu \otimes \left(1+\frac{u}{\kappa}P\right) +\left(1-\frac{1-u}{\kappa}P\right) \otimes p_\mu}{1\otimes 1+u(1-u)\frac{1}{\kappa^2}P\otimes P},\\
&& \Delta^{\mathcal{F}_{L,u}} (D)= \mathcal{F}_{L,u}\,\Delta D\, \mathcal{F}_{L,u}^{-1} = \left( D\otimes \frac{1}{1+\frac{u}{\kappa}P} +\frac{1}{1-\frac{1-u}{\kappa}P}\otimes D\right) \left( 1\otimes 1 +u(1-u)\frac{1}{\kappa^2}P\otimes P\right),\nonumber\\
&& \\
&& S^{\mathcal{F}_{L,u}}(p_\mu)=\frac{p_\mu }{1-(1-2u)\frac{1}{\kappa}P},\\
&& S^{\mathcal{F}_{L,u}}(D)=-\left( \frac{1-(1-2u)\frac{P}{\kappa}}{1+\frac{u}{\kappa}P}\right) D\left(1+\frac{u}{\kappa}P\right) ,
\end{eqnarray}
where $p_\mu,\, \mu\in \{0,1,...,n-1\}$ are momenta in the Minkowski spacetime and $P=v^\alpha p_\alpha$, where $v_\alpha v^\alpha\in \{1,0,-1\}$.

For $u=0$, $\mathcal{F}_{L,u=0}$ reduces to the Jordanian twist
\begin{eqnarray}
\mathcal{F}_{0}&=&\text{exp}(-\text{ln}(1-\frac{1}{\kappa}P)\otimes D)\nonumber \\
&=& \sum_{k=0}^\infty \left(-\frac{P}{\kappa}\right)^k \otimes \binom{-D}{k} = \text{exp}\left(-\text{ln}\left(1-\frac{1}{\kappa}P\right)\otimes D\right).
\end{eqnarray}
For $u=1/2$, $\mathcal{F}_{L,u=1/2}$ corresponds to the twist proposed in Refs. \cite{Tolstoy,Tolstoy2}. For $u=1$, it follows from Refs. \cite{cobtw} and \cite{MMss} that $\mathcal{F}_{L,u=1}$ is identical to the Jordanian twist
\begin{eqnarray}
\mathcal{F}_1&=&\text{exp}(-D\otimes \text{ln}(1+\frac{1}{\kappa}P))\nonumber \\
&=& \sum_{l=0}^\infty \binom{-D}{l} \otimes \left(\frac{P}{\kappa}\right)^l =\text{exp}\left(-D\otimes \text{ln}\left(1+\frac{1}{\kappa}\right)\right).
\end{eqnarray}
Hence, the family of twists $\mathcal{F}_{L,u}$ interpolates between twists $\mathcal{F}_0$ and $\mathcal{F}_1$.

Note that from $[P,D]= P$ it follows that $DP=P(D-1)$, and generally
\begin{eqnarray}
&& f(D)P= Pf(D-1), \\
&& f(D)P^m= P^m f(D-m).
\end{eqnarray}

Another family of twists induced with a 1-cochain $\omega_R=\text{exp}\left(-\frac{u}{\kappa}PD\right)$ is \cite{cobtw} 
\begin{equation}\label{FRu}
\mathcal{F}_{R,u}=\text{e}^{\frac{u}{\kappa}(PD\otimes 1+1\otimes PD)}\, \text{e}^{-\text{ln}\left(1-\frac{1}{\kappa}P\right)\otimes D}\, \text{e}^{-\Delta\left(\frac{u}{\kappa} PD\right)}.
\end{equation}
These twists satisfy the normalization and cocycle condition.

The corresponding deformed Hopf algebra is given by
\begin{eqnarray}
&& \Delta^{\mathcal{F}_{R,u}}(p_\mu) =\mathcal{F}_{R,u}\, \Delta p_\mu\, \mathcal{F}_{R,u}^{-1} =\frac{p_\mu \otimes \left( 1+\frac{u}{\kappa}P\right) +\left( 1-\frac{1-u}{\kappa}P\right) \otimes p_\mu}{1\otimes 1+\frac{u(1-u)}{\kappa^2}P\otimes P}, \\
&& \Delta^{\mathcal{F}_{R,u}} (D)=\mathcal{F}_{R,u}\, \Delta D\, \mathcal{F}_{R,u}^{-1}=\left( 1\otimes 1+\frac{u(1-u)}{\kappa^2} P\otimes P\right) \left( D\otimes \frac{1}{1+\frac{u}{\kappa}P} +\frac{1}{1-\frac{1-u}{\kappa} P} \right),\nonumber \\
  && \\
&& S^{\mathcal{F}_{R,u}} (p_\mu) =-\frac{p_\mu}{1-\frac{1-2u}{\kappa}P}, \\
&& S^{\mathcal{F}_{R,u}} (D)= -\left( 1-(1-u)\frac{P}{\kappa}\right) D \left(\frac{1-(1-2u)\frac{P}{\kappa}}{1-(1-u)\frac{P}{\kappa}}\right).
\end{eqnarray}
For $u=0$, $\mathcal{F}_{R,u=0}$ reduces to the Jordanian twist $\mathcal{F}_0$ and for $u=1$ it was shown in Ref. \cite{MMss} that $\mathcal{F}_{R,u=1}=\mathcal{F}_1$. Hence, the family of twists $\mathcal{F}_{R,u}$ interpolates between two Jordanian twists $\mathcal{F}_0$ and $\mathcal{F}_1$.

We point out that for $u=1/2$, $\mathcal{F}_{R,u=1/2}^{-1}=\mathcal{F}_{GZ}^{-1}$, where $\mathcal{F}_{GZ}^{-1}$ is the twist proposed in Ref. \cite{GZ}, theorem 2.20. Hence, the twist $\mathcal{F}_{R,u=1/2}$ is given by \eqref{FRu} and satisfies the normalization and cocycle condition automatically, as it is obtained as the cochain twist of a normalized 2-cocycle. It was shown in Ref. \cite{MMss} that the twist $\mathcal{F}^{-1}_{R,u}$ can be written as
\begin{equation}
\mathcal{F}_{R,u}^{-1}=\sum_{k,l=0}^\infty (u-1)^k\left(\frac{P}{\kappa}\right)^k\binom{D}{l} \otimes \left(\frac{u}{\kappa}P\right)^l\binom{D}{k}.
\end{equation}

Starting with
\begin{equation}
\mathcal{F}_{R,u}^{-1}= \text{e}^{\Delta(\frac{u}{\kappa}P D)}\, \text{e}^{-\Delta(\frac{u}{\kappa} DP)}\, \mathcal{F}_{L,u}^{-1} \,\text{e}^{\frac{u}{\kappa}(DP\otimes 1+1\otimes DP)}\, \text{e}^{-\frac{u}{\kappa}(PD\otimes 1+1\otimes PD)},
\end{equation}
and using (see also Eq. (39) in Ref.~\cite{cobtw})
\begin{equation}\label{e39u12}
\text{exp}\left(\frac{u}{\kappa} PD\right)\,\text{exp}\left(-\frac{u}{\kappa} DP\right)=1+\frac{u}{\kappa}P,
\end{equation}
we find
\begin{equation}
\mathcal{F}_{R,u}^{-1} = \mathcal{F}_{L,u}^{-1} \frac{1}{1\otimes 1+\frac{u(1-u)}{\kappa^2} P\otimes P}.
\end{equation}
Note that for $u = 1$, $\mathcal{F}_{L,u=1}= \mathcal{F}_{R,u=1}$.

\section{Two-parameter family of Jordanian twists and relations between them}

More generally we can define a two parameter family of twists $\mathcal{F}_{w,u}$, with coboundary twist $\omega_{w,u}=\text{exp}\left(-\frac{u}{\kappa}(D+w)P\right)$
\begin{eqnarray}
&& \mathcal{F}_{w,u} = \text{exp}\left( \frac{u}{\kappa}((D+w)P\otimes 1+1\otimes (D+w)P)\right)\,\text{exp}\left( -\ln \left(1-\frac{P}{\kappa}\right)\otimes D\right)\,\text{exp}\left(-\Delta\left(\frac{u}{\kappa} (D+w)P\right)\right).\nonumber\\
&& \label{Fwu}
\end{eqnarray}
For $w=1$ it coincides with $\mathcal{F}_{R,u}$ and for $w=0$ with $\mathcal{F}_{L,u}$. Let us use relation \eqref{e39u12}. For $u=1$, it yields
\begin{equation}
\text{e}^{\frac{P}{\kappa}D}\, \text{e}^{-D\frac{P}{\kappa}}= \text{e}^{\frac{1}{\kappa}(DP+P)}\text{e}^{-D\frac{P}{\kappa}}=\text{e}^{\text{ln}\left(1+\frac{P}{\kappa}\right)}.
\end{equation}
Generalizing this relation for arbitrary $w$ and using the BCH formula, it follows
\begin{equation}
\text{exp}\left(D\frac{P}{\kappa}+w\frac{P}{\kappa}\right) \, \text{exp}\left( -D\frac{P}{\kappa}\right)= \text{exp}\left( w\text{ln}\left(1+\frac{P}{\kappa}\right)\right) =\left(1+\frac{P}{\kappa}\right)^{w}.
\end{equation}
After rescaling $P\to uP$, $D\to D$, $w\to w$, it follows
\begin{equation}
\text{exp}\left( \frac{u}{\kappa}( D+w)P \right)\, \text{exp}\left(-\frac{u}{\kappa}DP\right) =\left( 1+\frac{u}{\kappa}P\right)^{w}.
\end{equation}
Using the above identity, the relation between $\mathcal{F}_{w,u}$ and $\mathcal{F}_{w=0,u}=\mathcal{F}_{L,u}$, one gets
\begin{eqnarray}
&& \mathcal{F}_{w,u}^{-1}=\Delta \left(1+\frac{u}{\kappa}P\right)^{w} \mathcal{F}_{L,u}^{-1} \frac{1}{\left(1\otimes 1+\frac{u}{\kappa}P\otimes 1\right)^{w}\left(1\otimes 1+1\otimes \frac{u}{\kappa}P\right)^{w}},\\
&& \mathcal{F}_{L,u}\mathcal{F}_{w,u}^{-1}= \Delta^{\mathcal{F}_{L,u}} \left(1+\frac{u}{\kappa}P\right)^{w} \left(\frac{1}{1\otimes 1+\frac{u}{\kappa}P\otimes 1}\right)^{w} \left(\frac{1}{1\otimes 1+1\otimes \frac{u}{\kappa}P}\right)^{w}\nonumber \\
&& \quad =\left( \frac{1}{1\otimes 1+\frac{u(1-u)}{\kappa^2}P\otimes P}\right)^w.
\end{eqnarray}
Hence,
\begin{equation}
\mathcal{F}_{w,u}^{-1}=\mathcal{F}_{L,u}^{-1}\left( \frac{1}{1\otimes 1+\frac{u(1-u)}{\kappa^2}P\otimes P}\right)^{w}.
\end{equation}
The same relations follow from the definition of the star product and the methods in Refs. \cite{Mercati,nas1608}. In the limit $u\to 0$, $\mathcal{F}_{w,u=0}^{-1}=\mathcal{F}_0^{-1}$,
and for $u=1$
\begin{equation}
\mathcal{F}^{-1}_{w,u=1}=\mathcal{F}^{-1}_1,\quad \forall w\in \mathbb{R}.\label{Fwuje1}
\end{equation}
Hence, the above one parameter family of twists $\mathcal{F}_{w,u}$ is identically equal to the simple Jordanian twist $\mathcal{F}_1$, generalizing the results from Sections 2 and 3. Note that for $u=1/2$ twists $F_{w,u=1/2}$ are $r$-symmetric in the first order in $1/\kappa$ for all values of $w$.

The quantum $R$-matrices are
\begin{equation}
R_{w,u}=\left(1\otimes 1+\frac{u(1-u)}{\kappa^2}P\otimes P\right)^w R_{L,u} \left(1\otimes 1+\frac{u(1-u)}{\kappa^2}P\otimes P\right)^{-w}.
\end{equation}
Note that the classical $r$-matrix does not depend on the parameters $w$ and $u$. 
\begin{equation}
r = \frac{1}{\kappa} (D \otimes P - P \otimes D).
\end{equation}

\section{Twisted coalgebra, star product and realizations}
% section 5
\label{sec:Hopfalg}

\subsection{Twisted coalgebra sector}
  
The corresponding Hopf algebra is defined with
\begin{eqnarray}
  \label{eq:Deltap}
&& \Delta^{\mathcal{F}_{w,u}}(p_\mu) =\Delta^{\mathcal{F}_{L,u}}p_\mu ,\\
\label{eq:Sp}
&& S^{\mathcal{F}_{w,u}}(p_\mu) =S^{\mathcal{F}_{L,u}}(p_\mu), \: \text{for arbitrary}\, w,\\
\label{eq:DeltaD}
&& \Delta^{\mathcal{F}_{w,u}}(D) =\left( 1\otimes 1+\frac{u(1-u)}{\kappa^2}P\otimes P\right)^{w} \Delta^{\mathcal{F}_{L,u}}(D) \left(1\otimes 1+\frac{u(1-u)}{\kappa^2}P\otimes P\right)^{-w},\\
\label{eq:SD}
&& S^{\mathcal{F}_{w,u}}(D)=-\left( \frac{1-\frac{1-2u}{\kappa}P}{1+\frac{u}{\kappa}P}\right)^{1-w} \left( 1-\frac{1-u}{\kappa}P\right)^wD \left( \frac{1-\frac{1-2u}{\kappa}P}{1-\frac{1-u}{\kappa}P}\right)^w \left(1+\frac{u}{\kappa}P\right)^{1-w}.
%\label{SFwuD}
\end{eqnarray}
Note that
\begin{eqnarray}
&& S^{\mathcal{F}_{w,u=0}}(D)=-\left(1-\frac{P}{\kappa}\right)D, \quad \forall w\\
&& S^{\mathcal{F}_{w,u=1}}(D)=-D\left(1+\frac{P}{\kappa}\right), \quad \forall w.
\end{eqnarray}

\subsection{Coordinate realizations of the $\kappa$-Minkowski spacetime and 
star product}

Let us define the Heisenberg(-Weyl) algebra generated
by commutative coordinates $x_\mu$ and the corresponding momenta $p_\mu$, satisfying
\begin{equation}
[x_\mu,x_\nu] = 0,\,\,\,\,\,\,
[p_\mu,p_\nu] = 0,\,\,\,\,\,\,
[p_\mu,x_\nu] = - i\eta_{\mu\nu}.
\end{equation}
The Heisenberg algebra acts on the space of functions
$f(x) = f(\{x_\mu\})$ of the commutative coordinates, where $x_\mu$ act
by multiplication and the action of generators $p_\mu$
and dilatation operator $D = i x_\mu p_\mu$ is given by 
\begin{equation}\label{eq:pact}
(p_\mu\triangleright f)(x) = -i\frac{\partial f(x)}{\partial x_\mu},
\end{equation}
\begin{equation}\label{eq:Dact}
(D\triangleright f)(x) = x_\mu \frac{\partial f(x)}{\partial x_\mu}.
\end{equation}
The subalgebra of coordinates becomes noncommutative due to a twist deformation,
replacing the usual multiplication with star product multiplication. This star
product is associative, because the twist $\mathcal{F}_{w,u}$ satisfies the 2-cocycle
condition. When the functions are exponentials $e^{ikx}$ and $e^{iqx}$,
we find their star product 
\begin{equation}\label{eq:starprod}
e^{ikx} * e^{iqx} = e^{i \mathcal{D}_\mu(k,q) x_\mu} G(k,q),
\end{equation}
  where
\begin{equation}\label{Dkq}
  \mathcal{D}_\mu(k,q) =
  \frac{k_\mu \left(1+\frac{u}{\kappa} (v\cdot q)\right)
    + \left(1-\frac{1-u}{\kappa}(v\cdot k)\right) q_\mu
    }{1+\frac{u(1-u)}{\kappa^2}(v\cdot k)(v\cdot q)},
\end{equation}
\begin{equation}\label{Gkq}
G(k,q) = \left(1+\frac{u(1-u)}{\kappa^2}(v\cdot k)(v\cdot q)\right)^{-w},
\end{equation}
and $k x=k_\alpha x_\alpha$, $q x=q_\alpha x_\alpha$, $v\cdot k = v_\alpha k_\alpha$, $v\cdot q = v_\alpha q_\alpha$. Results \eqref{eq:starprod}, \eqref{Dkq}, \eqref{Gkq} follow also from the methods in Refs. \cite{Mercati,nas1608}.

Directly from the twist $\mathcal{F}_{w,u}$, we also calculate the realizations of the noncommutative coordinates $\hat{x}_\mu$ in terms of the Heisenberg algebra generators,
\begin{equation}\label{eq:hxmu}
\hat{x}_\mu = m\left[\mathcal{F}_{w,u}^{-1}(\triangleright \otimes 1)(x_\mu\otimes 1)\right] = \left( x_\mu + i v_\mu\frac{1-u}{\kappa}D\right)\left(1+\frac{u}{\kappa} P\right) + i w\frac{u(1-u)}{\kappa^2}v_\mu P,
\end{equation}
where $m$ is the multiplication map $m:a\otimes b\mapsto a b$ of the Heisenberg algebra. These realizations are generalizations of those discussed in Refs.~\cite{govindarajan,stojic2,mkj,mkj2,mmss2}.

In the case $u = 0$, $\hat{x}_\mu = x_\mu + \frac{i}{\kappa} v_\mu D$ and in the
case $u = 1$, $\hat{x}_\mu = x_\mu\left(1+\frac{1}{\kappa}P\right)$.
Note that for $u = 0$ and $u = 1$ realizations do not depend on $w$. Noncommutative coordinates $\hat{x}_\mu$, \eqref{eq:hxmu}, generate the $\kappa$-Minkowski spacetime and satisfy
\begin{equation}
[\hat{x}_\mu,\hat{x}_\nu] = \frac{i}{\kappa}\left( v_\mu\hat{x}_\nu - v_\nu\hat{x}_\mu\right),
\end{equation}
\begin{equation}
[p_\mu,\hat{x}_\nu] = \left(-i\eta^\nu_\mu+\frac{i}{\kappa} v_\nu \frac{1-u}{\kappa}p_\mu\right)\left(1+\frac{u}{\kappa}P\right).
\end{equation}
Note that the term in $\hat{x}_\mu$ proportional to $w$ does not influence the commutation relations for $[\hat{x}_\mu,\hat{x_\nu}]$ and $[p_\mu,\hat{x}_\nu]$. 

\subsection{Real forms of Jordanian deformations}

For physical applications it is important to address the question if the symmetry Hopf algebras can be endowed with (compatible) *-structures. Hopf *-algebras are Hopf algebras with an involution $a\mapsto a^*$ satisfying identities, needed to treat unitarity and hermiticity in physical applications. In our case, to construct the *-involution, we start with $P^*=P$, $D^*=-D$ and find
\begin{eqnarray}
&& (\mathcal{F}_{w,u})^{*\otimes *} =\mathcal{F}_{w,u}^{-1} \left(1\otimes 1+\frac{u(1-u)}{\kappa^2}P\otimes P\right)^{2w-1}\: \text{and}\\
&& \left( S^{\mathcal{F}_{w,u}}(g)\right)^* =S^{\mathcal{F}_{(1-w),(1-u)}}(g^*)\vert_{-\kappa}, \: \forall g.
\end{eqnarray}
For $w=1/2$, it follows
\begin{eqnarray}
&& (\mathcal{F}_{1/2,u})^{*\otimes *} =\mathcal{F}_{1/2,u}^{-1},\\
&& \left(S^{\mathcal{F}_{1/2,u}}(D)\right)^* =S^{\mathcal{F}_{1/2,(1-u)}}(D^*)\vert_{-\kappa}.
\end{eqnarray}
Hence the corresponding twist is unitary for $w=1/2$, $u\in \mathbb{R}$ and also for $u=0$, $u=1$ and for arbitrary $w$.

Generally, the twist $\mathcal{F}_{w,u}$ is not unitary. If one wants to obtain a Hopf *-algebra structure on the deformation, there is a construction~\cite{majid} provided the condition
\begin{equation}\label{Mcond}
(S\otimes S)(\mathcal{F})^{*\otimes *}=\mathcal{F}^\tau,
\end{equation}
is satisfied, where $\mathcal{F}^\tau$ denotes the flipped (transposed) twist $\mathcal{F}$. In our case, for $\mathcal{F}_{w,u}$, we check 
\begin{equation}
(S\otimes S)(\mathcal{F}_{w,u})^{*\otimes *}= \mathcal{F}_{w,u}\vert_{-\kappa}=\mathcal{F}_{w,(1-u)}^\tau,
\end{equation}
and the condition \eqref{Mcond} is satisfied for $u=1/2$. A whole discussion of real forms of Jordanian deformations can be found in Section 4 of Ref. \cite{cobtw}.
%The generalized star operation *' is then defined with
%\begin{equation}
%g^{*'}=-S^{\mathcal{F}_{1/2,w}}(g^{*'}).\quad ?
%\end{equation}
%I.e., $P^{*'}=P$ ?, $D^{*'}=S_{1/2,w}(D)$ ?.

\section{Similarity transformations, the Poincar\'e-Weyl algebra and dispersion relations}

Let us consider the twist $\mathcal{F}_{w,u}$, \eqref{Fwu}, for $u=0$ and denote the corresponding generators with $P^0,\,D^0$, i.e.
\begin{equation}
\mathcal{F}_0=\text{exp}\left(-\ln \left(1-\frac{P_0}{\kappa}\right)\otimes D_0\right),
\end{equation}
leading to the following Hopf algebra
\begin{eqnarray}
&& \Delta^{\mathcal{F}_0}(p_\mu^0)=p^0_\mu \otimes 1+\left(1-\frac{P_0}{\kappa}\right)\otimes p^0_\mu, \\
&& \Delta^{\mathcal{F}_0}(D_0)=D_0\otimes 1 +\frac{1}{1-\frac{1}{\kappa}P_0}\otimes D_0,\\
&& S^{\mathcal{F}_0}(p^0_\mu)=-\frac{p^0_\mu}{1-\frac{1}{\kappa}P_0},\\
&& S^{\mathcal{F}_0}(D_0)=-\left(1-\frac{1}{\kappa}P_0\right)D_0.
\end{eqnarray}
We define Lorentz generators $M^0_{\mu\nu}$ generating the Poincar\'e-Weyl algebra together with $p^0_\mu$ and
$D^0$
\begin{eqnarray*}
&& [M^0_{\mu\nu},M^0_{\rho\sigma}] =
  -(\eta_{\mu\rho} M^0_{\nu\sigma} -\eta_{\mu\sigma}M^0_{\nu\sigma}
  +\eta_{\nu\rho}M^0_{\mu\sigma}-\eta_{\nu\sigma}M^0_{\mu\rho}),
\\
&&  [M^0_{\mu\nu},p^0_\lambda] = -(\eta_{\mu\lambda}p^0_\nu -\eta_{\nu\lambda}p^0_\lambda),
\\
&&  [D^0,M^0_{\mu\nu}] = [p^0_\mu,p^0_\nu] = 0,
\\
&&  [D^0,p^0_\mu] = - p^0_\mu.
\end{eqnarray*}
Then the coproduct and antipodes are
\begin{eqnarray}
\Delta^{\mathcal{F}_0}(M_{\mu\nu}^0) = M^0_{\mu\nu}\otimes 1 + 1\otimes M^0_{\mu\nu}
+ \frac{1}{\kappa}\frac{v_\mu p^0_\nu - v_\mu p^0_\mu}{1-\frac{1}{\kappa}p^0}\otimes D^0,\\
S^{\mathcal{F}_0}(M^0_{\mu\nu}) = - M^0_{\mu\nu}+\frac{1}{\kappa}(v_\mu p^0_\nu - v_\nu p^0_\mu) D^0.
\end{eqnarray}

Generalizing Ref.~\cite{mmmsprd}, and using similarity transformations induced by $\omega_{w,u}$, we obtain new generators
\begin{equation}\label{eq:pp0}
p_\mu =\exp\left(-\frac{u}{\kappa}(D_0+w)P_0\right)p_\mu^0\,\exp\left(\frac{u}{\kappa}(D_0+w)P_0\right)= \frac{p^0_\mu}{1-\frac{u}{\kappa}P_0},
    \end{equation}
   \begin{equation}\label{eq:DD0}
D = \exp\left(-\frac{u}{\kappa}(D_0+w)P_0\right)D_0\exp\left( \frac{u}{\kappa} (D_0+w)P_0\right) =D_0\left(1-\frac{u}{\kappa}P_0\right) -\frac{u}{\kappa}w P_0,
\end{equation}
 \begin{equation}\label{eq:MM0}
M_{\mu\nu} = \exp\left(-\frac{u}{\kappa}(D_0+w)P_0\right) M^0_{\mu\nu} \exp\left(\frac{u}{\kappa}(D_0+w)P_0\right) = M^0_{\mu\nu} - \frac{u}{\kappa}(D^0+w)(v_\mu p^0_\nu - v_\nu p^0_\mu),
\end{equation}
and the inverse relations
\begin{eqnarray}\label{eq:p0p}
&& p^0_\mu =\exp\left(\frac{u}{\kappa}(D_0+w)P_0\right)p_\mu \exp\left(-\frac{u}{\kappa}(D_0+w)P_0\right)= \frac{p_\mu }{1+\frac{u}{\kappa}P},\\ \label{eq:D0D}
&& D^0=\exp\left(\frac{u}{\kappa}(D_0+w)P_0\right) D\exp\left(-\frac{u}{\kappa}(D_0+w)P_0\right)=D\left(1+\frac{u}{\kappa}P\right) +\frac{u}{\kappa}w P,
\end{eqnarray}
\begin{equation}\label{eq:M0M}
M_{\mu\nu}^0 = \exp\left(\frac{u}{\kappa}(D_0+w)P_0\right) M_{\mu\nu} \exp\left(-\frac{u}{\kappa}(D_0+w)P_0\right) = M_{\mu\nu} + \frac{u}{\kappa}(D+w)(v_\mu p_\nu - v_\nu p_\mu).
\end{equation}
Note that
\begin{equation}
(D+w)P=(D_0+w)P_0.
\end{equation}
Using~(\ref{eq:pp0}),(\ref{eq:DD0}),(\ref{eq:MM0}),
we express $M_{\mu\nu},D,p_\mu,$ in terms of $M^0_{\mu\nu},D_0,p_\mu^0,$
and calculate the deformed coproducts of
$M_{\mu\nu},D,p_\mu$ in terms of $M^0_{\mu\nu}, D_0$ and $p^0_\mu$,
and then we use~(\ref{eq:p0p}),(\ref{eq:D0D}),(\ref{eq:M0M}) to reexpress
the resulting coproducts in terms of $M_{\mu\nu}$, $D$ and $p_\mu$, 
obtaining
\begin{eqnarray}
&& \Delta^{\mathcal{F}_{w,u}}(p_\mu)=\Delta^{\mathcal{F}_0}\left( \frac{p^0_\mu }{1-\frac{u}{\kappa}P_0} \right)=\frac{p_\mu \otimes \left(1+\frac{u}{\kappa}P\right) +\left(1-\frac{1-u}{\kappa}P\right)\otimes p_\mu }{1\otimes 1+ \frac{u(1-u)}{\kappa^2}P\otimes P},\label{deltaFwupmu} \\
&& \Delta^{\mathcal{F}_{w,u}}(D)= \Delta^{\mathcal{F}_{0}}\left( D_0\left( 1-\frac{u}{\kappa}P_0\right) -\frac{u}{\kappa}wP_0\right)\nonumber \\
&&= \left( 1\otimes 1+\frac{u(1-u)}{\kappa^2}P\otimes P\right)^w \left( D\otimes \frac{1}{1+\frac{u}{\kappa}P} +\frac{1}{1-\frac{1-u}{\kappa}P}\otimes D\right) \left( 1\otimes 1+\frac{u(1-u)}{\kappa^2} P\otimes P\right)^{1-w}, \\
  &&  \Delta^{\mathcal{F}_{w,u}}(M_{\mu\nu}) = \Delta^{\mathcal{F}_0}\left(M^0_{\mu\nu} - \frac{u}{\kappa}(D^0+w)(v_\mu p^0_\nu - v_\nu p^0_\mu)\right)
  = M_{\mu\nu}\otimes 1 + 1\otimes M_{\mu\nu} - \nonumber\\
  && - \frac{1}{\kappa}\left[
    (u-1)\frac{v_\mu p_\nu - v_\nu p_\mu}{1 + \frac{u-1}{\kappa} P}\otimes
    \left( D(1+\frac{u}{\kappa}P)+\frac{u}{\kappa}w P\right)
    + u \left( D + (D + w)\frac{u-1}{\kappa} P\right)\otimes \frac{v_\mu p_\nu
      - v_\nu p_\mu}{1 + \frac{u}{\kappa} P}\right],\nonumber\\
      && \label{deltaFwuM}
\end{eqnarray}
and similarly for the antipode,
\begin{eqnarray}
&& S^{\mathcal{F}_{w,u}}(p_\mu) =S^{\mathcal{F}_0}\left( \frac{p_\mu^0}{1-\frac{u}{\kappa}P_0} \right) =-\frac{p_\mu }{1+\frac{2u-1}{\kappa}P},\\
&& S^{\mathcal{F}_{w,u}}(D)= S^{\mathcal{F}_{0}} \left( D_0\left(1-\frac{u}{\kappa} P_0\right) -\frac{u}{\kappa}wP_0 \right) \nonumber\\
  &&\quad =-\frac{1+\frac{2u-1}{\kappa}P}{1+\frac{u}{\kappa}P} D\left( 1+\frac{u}{\kappa}P \right) +w\frac{u(1-u)}{\kappa^2} P^2 \frac{2+\frac{2u-1}{\kappa}P}{\left(1+\frac{u}{\kappa}P\right) \left(1+\frac{u-1}{\kappa}P\right)},\label{SFwuD}
  \\
  && S^{\mathcal{F}_{w,u}}(M_{\mu\nu}) = S^{\mathcal{F}_0}\left(M^0_{\mu\nu} - \frac{u}{\kappa}(D^0+w)(v_\mu p^0_\nu - v_\nu p^0_\mu)\right) \nonumber\\
  && = - M_{\mu\nu} - \frac{u}{\kappa}(D+w)(v_\mu p_\nu - v_\nu p_\mu)
  - \frac{v_\mu p_\nu - v_\nu p_\mu}{\kappa}
  \left[
    (u-1) D + u \frac{(u-1)(w-1)\frac{p}{\kappa} - w}{1+\frac{u}{\kappa}P}
    \right]. \nonumber\\
    &&\label{SFwuM}
\end{eqnarray}
in agreement with the results~\eqref{eq:Deltap}-\eqref{eq:SD} in Section~\ref{sec:Hopfalg}. We point out that the above equations~\eqref{deltaFwupmu}-\eqref{SFwuM} for $u=1$ do not depend on the parameter $w$, in accordance with \eqref{Fwuje1}.

For the Poincar\'{e} algebra generated by Lorentz generators $M_{\mu\nu}^0$ and $p_\mu^0$, the corresponding quadratic Casimir is $(p^0)^2=(p^0)_\alpha (p^0)^{\alpha}$. It is also invariant under the deformed Poincar\'e algebra generated with $M^0_{\mu\nu}$ and $p_\mu$, where $p_\mu$ is related to $p^0_\mu$ \eqref{eq:pp0} and $p^0_\mu$ to $p_\mu$ \eqref{eq:p0p}. Note that $p^2=p^\alpha p_\alpha$ is  not invariant under $M^0_{\mu\nu}$. Hence
\begin{equation}\label{eq:disp1}
(p^0)^2= \frac{p^2}{\left(1+\frac{u}{\kappa}P\right)^2}.
\end{equation}
Momenta $p_\mu$ correspond to a realization of $\hat{x}_\mu$ in terms of $x_\mu$ and $p_\mu$ (\ref{eq:hxmu}). The corresponding dispersion relation is explicitly given
\begin{equation}\label{disprel1}
E^2-\vec{p}^2=m^2\left(1+\frac{u}{\kappa}E\right)^2,
\end{equation}
where $m$ is the mass of the particle, $E$ is the energy and $\vec{p}$ is the momentum. Note that $m^2$ is the value of the deformed mass Casimir.

On the other hand, the quadratic Casimir for the Poincar\'{e} algebra generated with $M_{\mu\nu}$ and $p_\mu$ is $p^2$. It is also invariant under the deformed Poincar\'e algebra generated with $M_{\mu\nu}$ and $p^0_\mu$, where $p^0_\mu$ is related to $p_\mu$ \eqref{eq:p0p} and $p_\mu$ to $p^0_\mu$ \eqref{eq:pp0}. Note that $(p^0)^2$ is not invariant under $M_{\mu\nu}$. Hence
\begin{equation}\label{eq:disp2}
p^2= \frac{(p^0)^2}{\left(1-\frac{u}{\kappa}P_0\right)^2}.
\end{equation}
Momenta $p^0_\mu$ correspond to a realization of $\hat{x}_\mu$ expressed in terms of $x^0_\mu, p^0_\mu$. The corresponding dispersion relation is explicitly given
\begin{equation}\label{disprel2}
E^2-\vec{p}^2=m^2\left( 1-\frac{u}{\kappa}E\right)^2.
\end{equation}
In the limit $\kappa \to \infty$ of Eqs. \eqref{disprel1} and \eqref{disprel2} the standard bilinear mass shell condition is obtained. These two families of dispersion relations are related to each other by inverse transformation, with a parameter change $u\mapsto -u$.

The addition of momenta $k_\mu \oplus q_\mu $ also depends on the parameter $u$
\begin{equation}\label{eq:addition}
k_\mu \oplus q_\mu =\frac{k_\mu \left( 1+\frac{u}{\kappa}v\cdot q\right) +\left( 1+\frac{u-1}{\kappa}v\cdot k\right)q_\mu }{1+\frac{u(1-u)}{\kappa^2}(v\cdot k)\,(v\cdot q)}.
\end{equation}
Note that dispersion relations~(\ref{eq:disp1}),(\ref{eq:disp2}) and addition of momenta (\ref{eq:addition})
do not depend on parameter $w$, whereas the realization of
$\hat{x}_\mu$~(\ref{eq:hxmu}), the star product~(\ref{eq:starprod}), $\Delta^{\mathcal{F}_{w,u}} (D)$~(\ref{eq:DeltaD}), $\Delta^{\mathcal{F}_{w,u}} (M_{\mu\nu})$ \eqref{deltaFwuM}, $S^{\mathcal{F}_{w,u}}(D)$~(\ref{eq:SD}) and $S^{\mathcal{F}_{w,u}} (M_{\mu\nu})$ \eqref{SFwuM} do depend on $w$.

\textbf{Concluding remarks}. The physical interpretation depends on the realization of the generators of the Poincar\'e-Weyl algebra~\cite{mmmsprd,Mercati,kmps}.
%ref Mignemi prd, Kovačević plb, Mercati plb.
Particularly, the spectrum of the relativistic hydrogen atom depending on the parameter $u$ was investigated in Ref.~\cite{mmmsprd}. Differences in realizations of NC coordinates could also lead to different physical predictions, see e.g. Ref.~\cite{borgupta}, where dispersion relations and the time delay parameter were investigated. Mathematically equivalent deformations, that are related to nonlinear changes of symmetry generators and linked with similarity maps, may lead to differences in the description of physical phenomena.

%The physical interpretation and the spectrum of the relativistic hydrogen atom depend on the realisation of the generators of the Poincar\'{e}-Weyl algebra and were investigated in \cite{mmmsprd}. Applications of Jordanian deformations have also been of interest in some recent literature.
%%%%%%%%%%%%%%%%%%%%%%%%%%%%%%%%%%%%%%%%%%%%%%%%%%%%%%%%%%%%%%

\section*{Acknowledgements} Z.\ \v{S}.\ has been partly supported by grant no. 18-00496S of the Czech Science Foundation.

\end{document}